\documentclass[twocolumn,twoside,slac_two]{revtex4}
\usepackage{graphicx}
\usepackage{fancyhdr}

\usepackage[justification=RaggedRight,labelfont=bf]{caption}
\usepackage{subfig}

\usepackage{amsmath} 

\def\email#1{{e-mail: \tt#1}}

\begin{document}

\title{Constraining the magnetic field in the parsec-scale jets of the
brightest {\em Fermi} blazars with multifrequency VLBI observations}

%

\author{K.V. Sokolovsky\footnote{Send offprint requests to K.V. Sokolovsky\\ \email{ksokolov@mpifr-bonn.mpg.de}}, Y.Y. Kovalev}
\affiliation{MPIfR, Bonn, Germany and ASC Lebedev, Moscow, Russia}
\author{A.P. Lobanov, T. Savolainen}
\affiliation{MPIfR, Bonn, Germany}
\author{A.B. Pushkarev} 
\affiliation{MPIfR, Bonn, Germany; Pulkovo Obs., St Petersburg, Russia; CrAO, Crimea, Ukraine}
\author{M. Kadler}
\affiliation{Dr. Remeis-Sternwarte \& ECAP, Bamberg, Germany; CRESST/NASA
GSFC and USRA, MD, USA}

\begin{abstract}
The spatially resolved broad-band spectroscopy with Very Long Baseline
Interferometry (VLBI) is one of the few methods
that can probe the physical conditions inside blazar jets.
We report on measurements of the magnetic field strength in parsec-scale radio
structures of selected bright {\em Fermi} blazars, based on fitting the synchrotron spectrum
to VLBA images made at seven frequencies in a 4.6--43.2~GHz range.
Upper limits of $B_{\bot} \le 10^{-2}$--$10^{2}$~G (observer's frame) could be placed on the magnetic
field strength in 13 sources. Hard radio spectra ($-0.5 \le \alpha \le +0.1$,
$S_{\nu} \sim \nu^{\alpha}$) observed above the synchrotron peak may
either be an indication of a hard energy spectrum of the emitting electron
population or result from a significant inhomogeneity of the emitting region.
\end{abstract}

\maketitle

\thispagestyle{fancy}


\section{Introduction}
Blazars are active galactic nuclei characterized by highly variable
non-thermal continuum emission across the electromagnetic spectrum. The
Large Area Telescope (LAT, \cite{2009ApJ...697.1071A}) on board the {\it Fermi} $\gamma$-ray observatory, launched
on 2008 June~11,
provides a wealth of new information about high-energy radiation of blazars. It is believed
that this high-energy emission is intimately connected with the emission at
lower energies (radio -- optical) through the process of inverse Compton
scattering. Very Long Baseline Interferometry (VLBI), with its
spectacular angular resolution, provides the most detailed view of inner
regions in blazars
\cite{1997ARA&A..35..607Z}. In most cases, VLBI observations reveal
one-sided relativistically beamed parsec-scale jets originating in a bright, compact
region called the ``core''.

The spatially resolved broad-band spectroscopy is one of the few methods
that can probe the physical conditions inside blazar jets. Despite the great
potential of this method, its applications are still relatively rare 
(see \cite{Savolainen} and references therein),
owing to difficulties in implementing it and interpreting results obtained \cite{1998A&AS..132..261L}.
In addition, multifrequency VLBI observations require
large amounts of observing time and most VLBI arrays lack technical
capability to perform simultaneous multifrequency observations.

We utilize the Very Long Baseline Array (VLBA, \cite{1994IAUS..158..117N}) which has a unique
capability to conduct observations quasi-simultaneously at many frequencies,
needed to constrain physical conditions in the regions where blazar radio emission
originates. We reconstruct radio spectra of parsec-scale features in jets of
selected $\gamma$-ray-bright blazars and compare them with the simple
homogeneous synchrotron source model, which allows us to derive information
about the magnetic field strength and particle energy distribution.
During the first year of the {\it Fermi} scientific operations, we used the VLBA to observe
20 blazars, that were identified prior to the launch of {\it Fermi} as expected bright $\gamma$-ray emitters. In
this paper, we present results for 17 of them. The data analysis for three
remaining sources is still ongoing.

\section{Observations and data reduction}
The observations were conducted simultaneously at seven frequencies in a
4.6--43.2~GHz range. After the initial calibration in the {\it AIPS} package
\cite{1981NRAON...3....3F},
the sources were self-calibrated and imaged independently at each frequency
using the {\it Difmap} software \cite{Shepherd}. Whenever feasible, the frequency-dependent core
position (the ``core shift'' effect, see for a detailed discussion
\cite{1998A&A...330...79L,2008A&A...483..759K}) was taken into account. 
Image alignment is a major problem
for spectral VLBI analysis because the self-calibration technique
used to obtain high-quality images
causes loss of information about the absolute position of a source.
Positions of optically thin jet features or the core itself 
(if the ``core shift'' effect is negligible for a given source) are
then used to align VLBI images obtained at different frequencies.

In order to reconstruct the spectrum in a given pixel of an image (``pixel-based
approach''), the sum 
of CLEAN components around this pixel was used. The CLEAN components were 
summed inside a radius $R = \Theta(\text{lowest frequency})/ 3$,
where $\Theta(\text{lowest frequency})$ is the half-power beam width
at the lowest observing frequency (4.6~GHz).
This approach has a major advantage over the direct use of flux densities from
reconstructed images because it avoids
artifacts introduced by convolution with a Gaussian beam while
preserving the angular resolution achieved.

\section{Results}
The radio spectra in the parsec-scale core regions (marked as shaded areas on
Fig.~\ref{fig:images_and_spectra}) of the observed blazars
have diverse properties. Most of them are nearly flat, with signs of 
curvature at lower frequencies, which we interpret as synchrotron
self-absorption. In one source (NRAO~530, Fig.~\ref{fig:1730m130s}), the spectral turnover is
particularly prominent and well sampled by our observations. Cores in three
sources, 0716$+$714 (Fig.~\ref{fig:0716p714s}), OJ~287
(Fig.~\ref{fig:0851p202s}), and 3C~454.3 (Fig.~\ref{fig:2251p158s}), show inverted spectra which are slightly
curved. The core spectrum of only one source (1510$-$089, Fig.~\ref{fig:1510m089s}) can be adequately described
by a simple power law in the whole 4.6--43.2~GHz frequency range.
In the following, we focus on sources that show a well defined
spectral turnover, and thus can be used to determine the physical
conditions in the emitting plasma.

We fit a homogeneous synchrotron source model \cite{Pacholczyk} (see
the Appendix for more details) to the observed spectra in order to obtain
the magnetic field strength and particle energy distribution. The sources for which the 
data reduction is complete are presented in Table~\ref{tab:one}.
For 13 sources, the model provided an adequate fit to the
observed spectra (Fig.~\ref{fig:images_and_spectra}) while four sources
exhibit flat or inverted spectra (Fig.~\ref{fig:bad_spectra}) that cannot be described by the
homogeneous synchrotron model.

\section{Discussion}

\subsection{Model reliability}
The physical parameters presented in Table~\ref{tab:one} correspond to
typical values for emitting electrons inside a large area of a few
milliarcseconds in size (shaded areas on Fig.~\ref{fig:images_and_spectra}). 
In most cases, only an
upper limit on the magnetic field strength could be placed, because most
regions where the spectral turnover is detected remain unresolved.
The results are consistent
with estimates obtained by a method based on model-fitting bright VLBI
components with 2D Gaussian components (see Fig.~\ref{fig:BLLac_spectra_comp}
for results on BL~Lacertae). The model-fitting based method is described in
detail in \cite{Savolainen} where it was applied to 3C~273 and the average
magnetic field strength $B_{\bot} \sim 10^{-2}$~G was obtained for the 8~GHz core
region which is comparable in size to the region probed in our research.
This value is in a good agreement with our upper limit at $B_{\bot} \le 0.016$~G
given in Table~\ref{tab:one}.

\begin{table*}[!ht]
 \caption{Physical parameters estimation}
 \begin{tabular}{c c c c r l}
 \hline
 IAU Name (B1950.0) & Alternative Name &   Epoch    & ~~$p^\mathrm{a,b}$~~ & $B_{\bot}^\mathrm{b,c}$ {\small [G]} & ~~Comments \\
              \hline
0235$+$164 & AO~0235$+$16      & 2008-09-02 & $0.8$ & $\le0.89$ & \\
0528$+$134 &                   & 2008-10-02 & $1.4$ & $\le0.12$ & \\
0716$+$714 &                   & 2009-02-05 &  &         & ~~inverted spectrum $\alpha=0.4$, model N/A$^\mathrm{d}$ \\
0827$+$243 & OJ~248            & 2009-04-09 & $1.6$ & $\le3.4$ & \\
0851$+$202 & OJ~287            & 2009-02-02 &  &         & ~~inverted spectrum $\alpha=0.7$, model N/A$^\mathrm{d}$ \\
1219$+$285 & W~Com             & 2009-05-14 & $0.8$ & $\le108$ & ~~preliminary analysis\\
1226$+$023 & 3C~273            & 2009-02-05 & $2.0$ & $\le0.016$ & \\
1253$-$055 & 3C~279            & 2009-02-02 & $1.4$ & $\le0.89$ & \\ 
1510$-$089 &                   & 2009-04-09 &  &        & ~~inverted spectrum $\alpha=0.2$, model N/A$^\mathrm{d}$ \\
1633$+$383 & 4C~38.41          & 2009-06-20 & $1.0$ & $\le0.24$ & ~~preliminary analysis\\
1652$+$398 & Mrk~501           & 2009-05-14 & $1.6$ & $\le3.5$ & ~~preliminary analysis\\ 
1730$-$130 & NRAO~530          & 2009-06-20 & $1.6$ & $\le3.6$ & ~~preliminary analysis\\ 
1959$+$650 &                   & 2008-10-23 & $1.4$ & $\le17$ & \\
2155$-$304 &                   & 2008-09-05 & $1.0$ & $\le26$ & \\
2200$+$420 & BL~Lac            & 2008-09-02 & $1.0$ & $\le0.44$ & \\
2251$+$158 & 3C~454.3          & 2008-10-02 &  &        & ~~inverted spectrum $\alpha=0.8$, model N/A$^\mathrm{d}$ \\
2344$+$514 &                   & 2008-10-23 & $1.2$ & $\le57$ & ~~43~GHz data not included in the fit\\ 
 \hline
 
 \multicolumn{6}{l}{$^\mathrm{a}$ $p$ is the power law index in the electron energy distribution $N(E)= N_0 E^{-p}$.}\\
 \multicolumn{6}{l}{Note, that for the optically thin part of the synchrotron spectrum $p = 1 - 2 \alpha$,}\\ 
 \multicolumn{6}{l}{where $\alpha$ is defined as $S_{\nu} \sim \nu^{\alpha}$.}\\
 \multicolumn{6}{l}{$^\mathrm{b}$ The estimates correspond to the region of parsec-scale radio core.}\\
 \multicolumn{6}{l}{$^\mathrm{c}$ The values are in the observer's frame.}\\
 \multicolumn{6}{l}{$^\mathrm{d}$ The homogeneous synchrotron source model is not applicable for this spectrum.}\\

\end{tabular} 
\label{tab:one}
\end{table*}

\subsection{Electron energy spectrum}
The spectra above the synchrotron turnover are nearly flat or slightly
inverted in core regions of all sources investigated in this work. This may result from blending of a few
emitting regions with different peak frequencies (an example of such
blending may be found
in the core region of BL Lacertae, see the top right panel in
Fig.~\ref{fig:BLLac_spectra_comp}, see also \cite{Savolainen} for 3C~273).
Alternatively, a nearly flat spectrum may imply a hard energy spectrum of the
emitting electrons (as may be the case for the component B1 in the jet of 
BL~Lacertae whose spectrum is presented in the lower right panel of
Fig.~\ref{fig:BLLac_spectra_comp}; see
also Table~\ref{tab:two}). The hard electron spectrum is difficult to explain by
the conventional first-order Fermi acceleration mechanism. However, the hard
spectrum can be produced by the second-order Fermi mechanism (``stochastic
acceleration'') \cite{2005ApJ...621..313V}.

\section{Summary and prospects}
We have placed upper limits of $B_{\bot} \le 10^{-2}$--$10^{2}$~G
(observer's frame) on the magnetic
field strength in 13 out of 17 {\it Fermi} blazars (Table~\ref{tab:one}).
Spectra of four blazars could not be described by the simple
homogeneous synchrotron source model, and no constraints on the magnetic field and
particle energy spectrum could be obtained for these sources.
Hard spectra ($-0.5 \le \alpha \le +0.1$, $S_{\nu} \sim \nu^{\alpha}$) observed
in the blazar cores above the synchrotron peak may
either indicate a hard energy spectrum of the relativistic electron  
population in the jet or result from significant inhomogeneity (of the
magnetic field and plasma parameters or just optical depth) across the emitting region.

Since $\gamma$-ray emission in blazars is suggested to originate from regions
spatially close to the VLBI core \cite{2009ApJ...696L..17K}, the estimates of the magnetic field
strength and electron energy distribution presented in Table~\ref{tab:one} could be used
to constrain broad-band Spectral Energy Distribution (SED) models. Simultaneously with the VLBA observations described
here, {\it Swift} X-ray/ultraviolet/optical target of opportunity observations
were performed. When combined with the {\it Fermi} $\gamma$-ray data this
will provide a unique data set for the SED modeling. Analysis of this data set is
presently underway.

\begin{acknowledgments}
K.~Sokolovsky is supported by the International Max-Planck Research
School (IMPRS) for Astronomy and Astrophysics at the universities of Bonn
and Cologne. T.~Savolainen is a research fellow of the
Alexander von Humboldt foundation. T.~Savolainen also acknowledges support by the
Academy of Finland grant 120516. Y.~Kovalev was supported by the return
fellowship of Alexander von Humboldt foundation and the Russian Foundation
for Basic Research (RFBR) grant 08-02-00545. This research has been
partially funded by the Bundesministerium f\"ur Wirtschaft und Technologie under 
Deutsches Zentrum f\"ur Luft- und Raumfahrt grant number 50OR0808.
This work is based on data obtained from the National Radio
Astronomy Observatory's Very Long Baseline Array (VLBA), project BK150. The National Radio
Astronomy Observatory is a facility of the National Science Foundation
operated under cooperative agreement by Associated Universities, Inc. 
This research has made use of NASA's Astrophysics Data System; the NASA/IPAC
Extragalactic Database (NED) which is operated by the JPL, Caltech, under contract with the
NASA, the SIMBAD database, operated at CDS, Strasbourg, France. The authors
are grateful to E.~Ros for reviewing this manuscript.
\end{acknowledgments}

\newpage

\section*{Appendix: synchrotron source model used in pixel-based approach}
Following \cite{Pacholczyk}, we consider a uniform cloud of relativistic electrons in the external
uniform magnetic field with component $B_{\bot}$ perpendicular to the line
of sight. We assume that the electron energy distribution is described by a power law:
$$N(E) = N_0 E^{-p} \text{,}$$
where $N_0$ and $p$ are constants.
For this case, we can write the radiation transfer equation
$$\frac{d I_{\nu}}{d s} = - \kappa_{\nu} I_{\nu} + j_{\nu} \text{,}$$
where $I_{\nu}$ is the specific intensity at the frequency $\nu$, $d s$ is the differential element
of path length, $j_{\nu}$ is the emission coefficient (emissivity), $\kappa_{\nu}$
is the absorption (extinction) coefficient.
If $j_{\nu}$ and $\kappa_{\nu}$ are constant along the line of sight, the
solution of the radiation transfer equation is
$$\label{eq:perenos_v_odnorodnoy_srede}
I_{\nu} = \int\limits_{0}^{L}
j_{\nu}e^{-\int\limits_{0}^{s}\kappa_{\nu}ds}ds =
\frac{j_{\nu}}{\kappa_{\nu}}(1 - e^{-\kappa_{\nu} L}) \text{,}$$
where $L$ is the size of the electron cloud along the line of sight, and $\kappa_{\nu} L
= \tau_{\nu}$ is the optical depth.

Following \cite{Pacholczyk}, this solution can be written in the form:
\begin{equation}
\label{eq:perenos_pah}
I_{\nu}=S(\nu_1)J(\frac{\nu}{\nu_1},p) \text{,}
\end{equation}
where
\begin{equation}
\label{eq:nu1_B}
\nu_1 = 2 c_1 (N_0 L c_6(p))^{2/(p+4)} B_{\bot}^{(p+2)/(p+4)} \text{,}
\end{equation}
\begin{equation}
\label{eq:Snu1_B}
S(\nu_1) = \frac{c_5(p)}{c_6(p)} B_{\bot}^{-1/2} \frac{\nu_1}{2 c_1}^{5/2} \text{,}
\end{equation}
$$J(z,p) = z^{5/2} (1 - \exp[-z^{-(p+4)/2}]) \text{,}$$
$$c_1 = \frac{3 e}{4 \pi m_e^3 c^5} = 6.27 \times 10^{18} \text{,}$$
$$c_3 = \frac{\sqrt{3}}{4 \pi} \frac{e^3}{m_e c^2} = 1.87 \times 10^{-23} \text{,}$$
$$c_5(p) = \frac{1}{4} c_3 \Gamma{\left(\frac{3 p - 1}{12}\right)} \Gamma{\left(\frac{3 p + 7}{12}\right)} \left(\frac{p + 7/3}{p + 1}\right) \text{,}$$
$$c_6(p) = \frac{1}{32} \left(\frac{c}{c_1}\right)^2 c_3 \left( p + \frac{10}{3} \right) \Gamma{\left(\frac{3 p + 2}{12}\right)} \Gamma{\left(\frac{3 p + 10}{12}\right)} \text{,}$$\\
where $e$ is the electron charge, $m_e$ is the electron mass, and $c$ is the
speed of light in vacuum.

The power law index $p$ of the electron energy distribution may be found
from the spectral index $\alpha$ ($S_{\nu} \sim \nu^{\alpha}$) in the optically thin part
of the spectrum: 
$p = 1 - 2 \alpha$.
The optical depth $\tau_{m}$ corresponding to the observed spectral peak
frequency $\nu_{m}$ may be found from equation $\exp[\tau_{m}] = 1 +
\frac{p+4}{5} \tau_{m}$.

The frequency $\nu_1$ corresponding to $\tau_{\nu_1} = 1$ may be found from
$\nu_{m}$, $p$ and $\tau_{m}$: $\nu_1 = \tau_{m}^{2/(p+4)} \nu_{m}$.

From equations (\ref{eq:perenos_pah}), (\ref{eq:nu1_B}) and
(\ref{eq:Snu1_B}),
we can derive the perpendicular component of the magnetic field:
\begin{equation}
B_{\bot} = \left(\frac{c_5(p)}{c_6(p)}\right)^2 S(\nu_1)^{-2} \left(\frac{\nu_1}{2 c_1}\right)^5 \text{.}
\end{equation}
Note, that if the size of an emitting region is smaller then the angular
resolution, only a {\it lower limit} on the intensity and, correspondingly,
an {\it upper limit} on the magnetic field $B_{\bot}$ can be derived.

\begin{figure*}[!hbt]
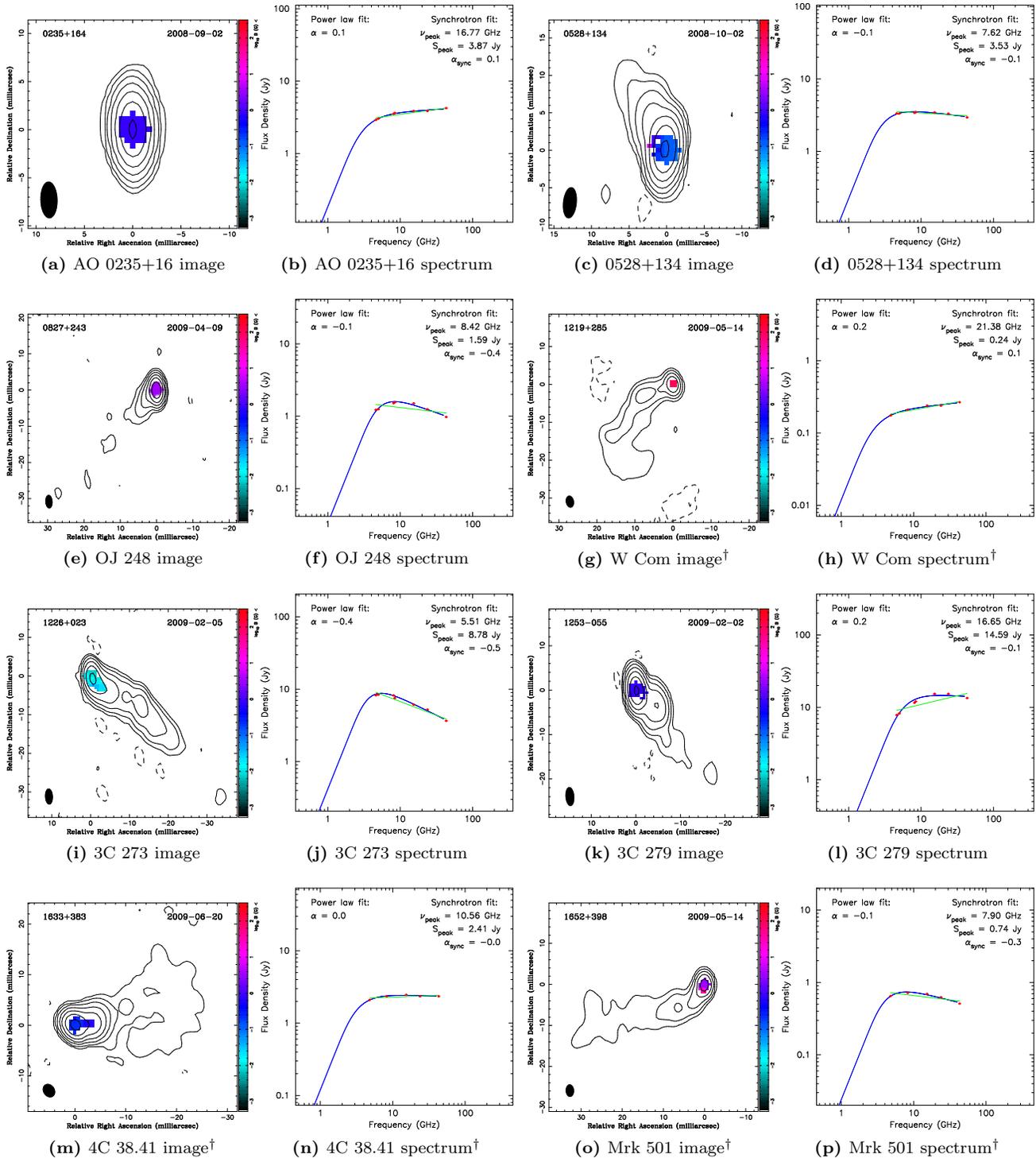

  \captionsetup{labelsep=space}
  \centering

  \subfloat[AO~0235$+$16 image]{\label{fig:0235p164}\includegraphics[width=0.25\textwidth]{0235+164.B_map.eps}}
  \subfloat[AO~0235$+$16 spectrum]{\label{fig:0235p164s}\includegraphics[width=0.25\textwidth]{0235+164.eps}}~~
  \subfloat[0528$+$134 image]{\label{fig:0528p134}\includegraphics[width=0.25\textwidth]{0528+134.B_map.eps}}
  \subfloat[0528$+$134 spectrum]{\label{fig:0528p134s}\includegraphics[width=0.25\textwidth]{0528+134.eps}}\\
  \subfloat[OJ~248 image]{\label{fig:0827p243}\includegraphics[width=0.25\textwidth]{0827+243.B_map.eps}}                
  \subfloat[OJ~248 spectrum]{\label{fig:0827p243s}\includegraphics[width=0.25\textwidth]{0827+243.eps}}~~
  \subfloat[W~Com image$^\dag$]{\label{fig:1219p285}\includegraphics[width=0.25\textwidth]{1219+285.B_map.eps}}
  \subfloat[W~Com spectrum$^\dag$]{\label{fig:1219p285s}\includegraphics[width=0.25\textwidth]{1219+285.eps}}\\
  \subfloat[3C~273 image]{\label{fig:1226p023}\includegraphics[width=0.25\textwidth]{1226+023.B_map.eps}}                
  \subfloat[3C~273 spectrum]{\label{fig:1226p023s}\includegraphics[width=0.25\textwidth]{1226+023.eps}}~~
  \subfloat[3C~279 image]{\label{fig:1253m055}\includegraphics[width=0.25\textwidth]{1253-055.B_map.eps}}
  \subfloat[3C~279 spectrum]{\label{fig:1253m055s}\includegraphics[width=0.25\textwidth]{1253-055.eps}}\\
  \subfloat[4C~38.41 image$^\dag$]{\label{fig:1633p383}\includegraphics[width=0.25\textwidth]{1633+383.B_map.eps}}
  \subfloat[4C~38.41 spectrum$^\dag$]{\label{fig:1633p383s}\includegraphics[width=0.25\textwidth]{1633+383.eps}}~~
  \subfloat[Mrk~501 image$^\dag$]{\label{fig:1652p398}\includegraphics[width=0.25\textwidth]{1652+398.B_map.eps}}                
  \subfloat[Mrk~501 spectrum$^\dag$]{\label{fig:1652p398s}\includegraphics[width=0.25\textwidth]{1652+398.eps}}\\

  \caption{VLBI images at 8.4~GHz and radio continuum spectra of sources for which the synchrotron
  self-absorption turnover frequency was successfully measured. The shaded area on the images
  marks the region where the synchrotron turnover was detected.
  The color of the shaded area represents the upper limit of the magnetic field strength ($B_{\bot}$).
  The values of $B_{\bot}$ and $p$ presented in Table~\ref{tab:one} correspond to this area.
  On the spectrum plots, the blue curves represent the homogeneous synchrotron source
  model, while the green lines represent a simple power law model. Best fit
  parameters of both models are shown on each plot. The $\dag$ symbol marks
  sources for which the analysis results are preliminary.}
  \label{fig:images_and_spectra}
  
\end{figure*}

\setcounter{figure}{0}

\begin{figure*}[h!]
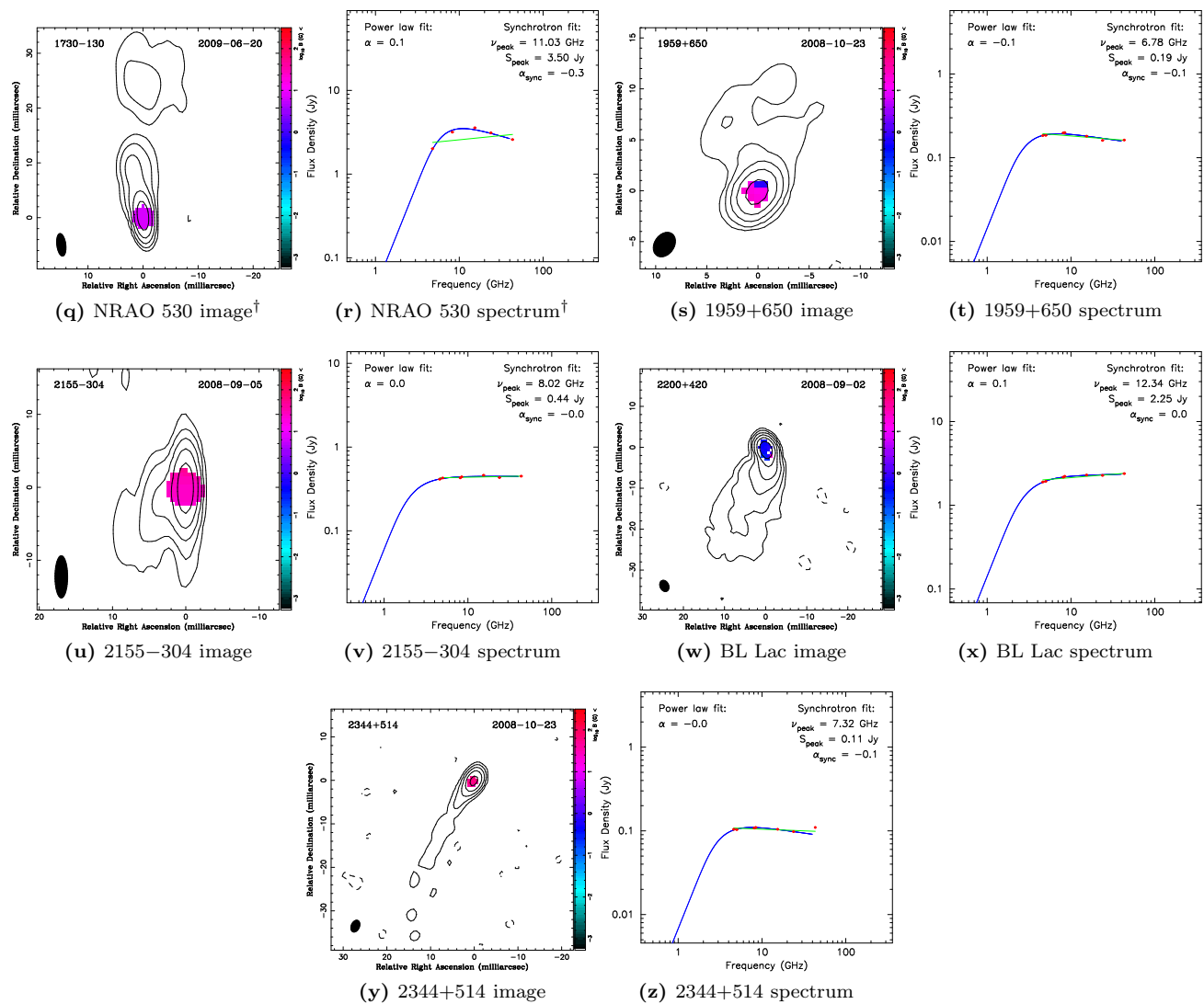

  \setcounter{subfigure}{16}
  \captionsetup{labelsep=space}
  \centering
  \subfloat[NRAO~530 image$^\dag$]{\label{fig:1730m130}\includegraphics[width=0.25\textwidth]{1730-130.B_map.eps}}
  \subfloat[NRAO~530 spectrum$^\dag$]{\label{fig:1730m130s}\includegraphics[width=0.25\textwidth]{1730-130.eps}}~~
  \subfloat[1959$+$650 image]{\label{fig:1959p650}\includegraphics[width=0.25\textwidth]{1959+650.B_map.eps}}                
  \subfloat[1959$+$650 spectrum]{\label{fig:1959p650s}\includegraphics[width=0.25\textwidth]{1959+650.eps}}\\
  \subfloat[2155$-$304 image]{\label{fig:2155m304}\includegraphics[width=0.25\textwidth]{2155-304.B_map.eps}}
  \subfloat[2155$-$304 spectrum]{\label{fig:2155m304s}\includegraphics[width=0.25\textwidth]{2155-304.eps}}~~
  \subfloat[BL~Lac image]{\label{fig:2200p420}\includegraphics[width=0.25\textwidth]{2200+420.B_map.eps}}                
  \subfloat[BL~Lac spectrum]{\label{fig:2200p420s}\includegraphics[width=0.25\textwidth]{2200+420.eps}}\\
  \subfloat[2344$+$514 image]{\label{fig:2344p514}\includegraphics[width=0.25\textwidth]{2344+514.B_map.eps}}
  \subfloat[2344$+$514 spectrum]{\label{fig:2344p514s}\includegraphics[width=0.25\textwidth]{2344+514.eps}}

  \caption{continued}
  \label{fig:images_and_spectra}
\end{figure*}

\begin{figure*}[h!]
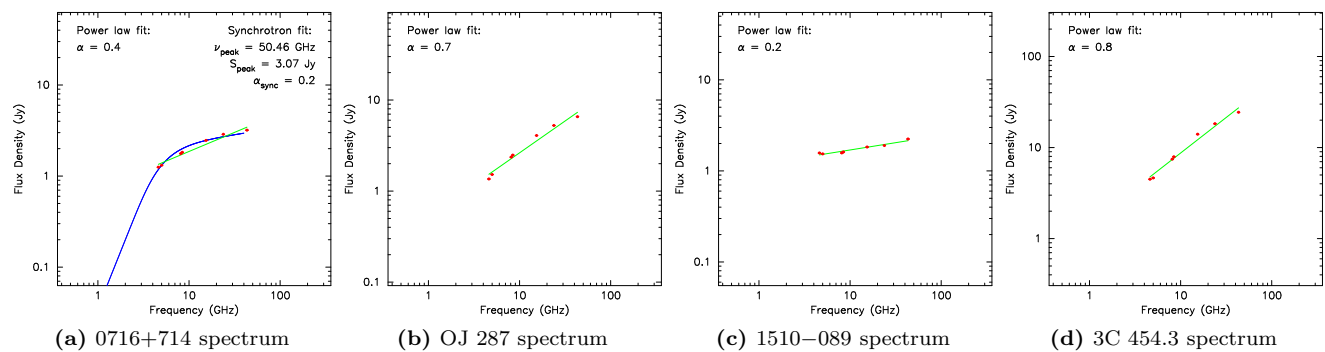

  \centering
  \captionsetup{labelsep=space}
  \subfloat[0716$+$714 spectrum]{\label{fig:0716p714s}\includegraphics[width=0.25\textwidth]{0716+714.eps}}~
  \subfloat[OJ~287 spectrum]{\label{fig:0851p202s}\includegraphics[width=0.25\textwidth]{0851+202.eps}}~
  \subfloat[1510$-$089 spectrum]{\label{fig:1510m089s}\includegraphics[width=0.25\textwidth]{1510-089.eps}}~
  \subfloat[3C~454.3 spectrum]{\label{fig:2251p158s}\includegraphics[width=0.25\textwidth]{2251+158.eps}}

  \caption{VLBA spectra of sources whose spectra cannot be adequately
described by the simple homogeneous model. The spectra correspond to
parsec-scale core regions. Green lines represent a power law fit. It is
possible to fit a homogeneous synchrotron source model (blue curve) to the observed
spectrum of 0716$+$714, however, a simple power law provides better (lower $\chi^2$) fit.}
  \label{fig:bad_spectra}
\end{figure*}

\begin{figure*}[t]
\centering
\captionsetup{labelsep=space}
\includegraphics[width=\textwidth]{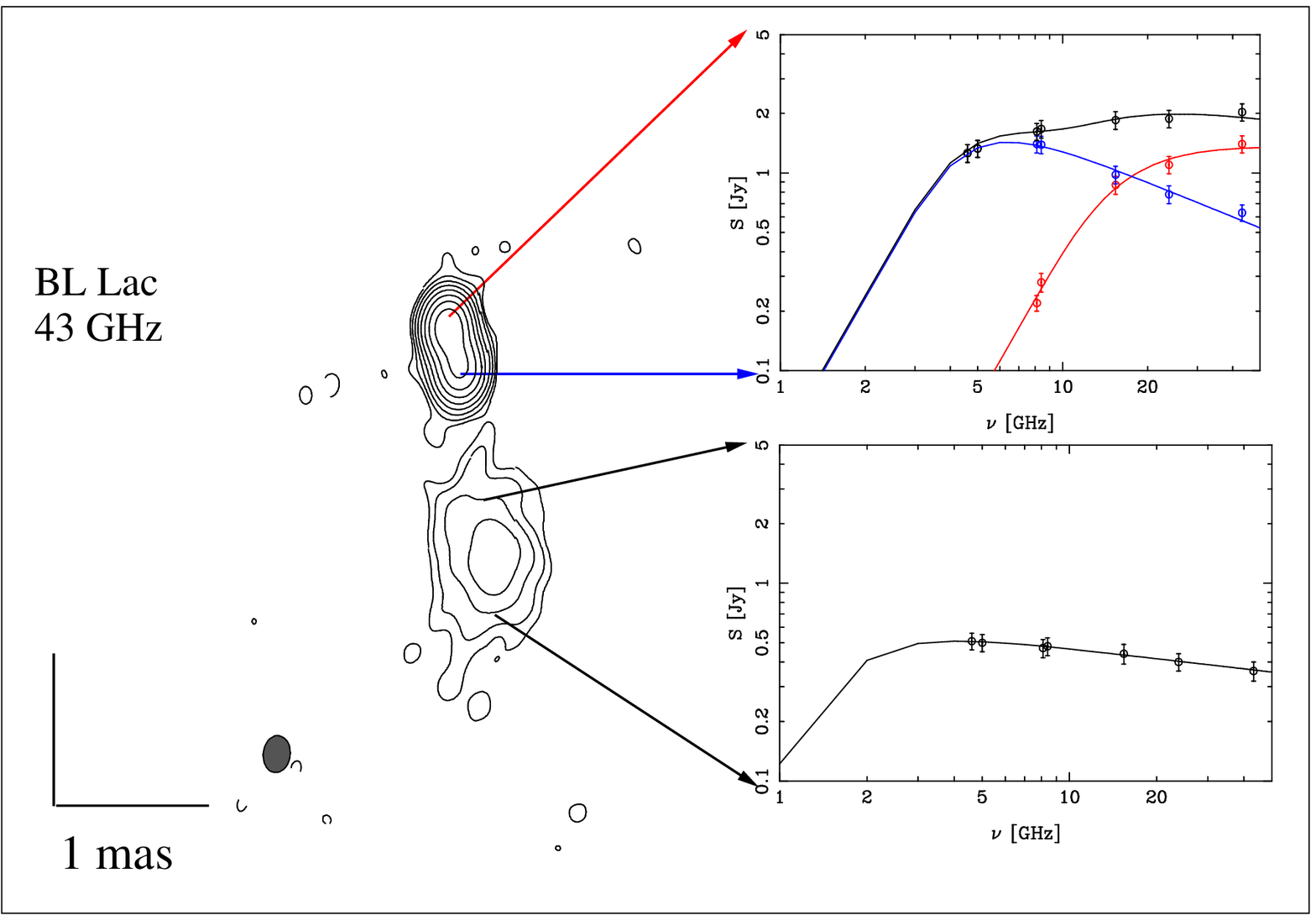}
\caption{Results of the model-fitting based analysis of BL Lacertae
(2200$+$420). The contours represent the CLEAN image at 43~GHz, the beam
size and the linear scale are indicated at the lower left panel.
The source structure was approximated with Gaussian components (see
Table~\ref{tab:two}). The fitting was conducted in the the $uv$-plane. 
The two inner components (C1 its position is marked by the red arrow, C2
marked by the blue arrow) are unresolved while the third component (B1
marked by two black arrows) is resolved at 43~GHz. 
Fig.~\ref{fig:2200p420s} presents the spectrum of this source obtained in
the course of pixel-based analysis. This spectrum corresponds to the
combined spectrum (black) of components C1 (red) and C2 (blue) presented in
the upper right panel of the current figure.
The peak frequency for B1 is not well constrained, however,
following \cite{Marscher} we can put an upper limit on the magnetic field
strength in this component $B<0.06$~G (observer's frame).}
\label{fig:BLLac_spectra_comp}
\end{figure*}

\begin{table*}[!b]
 \caption{BL Lacertae model component parameters}
 \centering
 \begin{tabular}{l c c r c}
 \hline
 Component~~~ &  ~~~Distance~~~ &  ~~~$\nu_m$~~~  &   ~~~$S_m$ &~~~  $\alpha$  \\
           &   {\small [mas]}   &  {\small [GHz]} &    {\small [Jy]}  &            \\     
 \hline
 C1 (core) &       $0$          &   $>43$          &   $>1.02$       & $\dots$    \\
 C2        &       $0.26$       &   $  6.4$        &   $ 1.43$       & $-0.58$ \\
 B1        &       $1.47$       &   $ <4.1$        &   $>0.51$       & $-0.17$ \\
 \hline
 
 \multicolumn{5}{l}{$\nu_m$ --- synchrotron self-absorption peak frequency.}\\   
 \multicolumn{5}{l}{$S_m$  --- synchrotron self-absorption peak flux density.}\\
 \multicolumn{5}{l}{$\alpha$ --- spectral index in the optically thin part of the spectrum.}\\
 \end{tabular}
 \label{tab:two}
\end{table*}


\begin{thebibliography}{99} 

\bibitem{2009ApJ...697.1071A} Atwood W.~B., et al., 2009, ApJ, 697, 1071 

\bibitem{1997ARA&A..35..607Z} Zensus J.~A., 1997, ARA\&A, 35, 607

\bibitem{Savolainen} Savolainen T., Wiik K., Valtaoja E., Tornikoski M. 2008, ASPC, 386, 451

\bibitem{1998A&AS..132..261L} Lobanov, A.~P.\ 1998, A\&AS,132, 261 

\bibitem{1994IAUS..158..117N} Napier P.~J., 1994, IAUS, 158, 117 

\bibitem{1981NRAON...3....3F} Fomalont E., 1981, NRAON, 3, 3 

\bibitem{Shepherd} Shepherd M.C., Pearson T.J. and Taylor G.B. 1994, BAAS, 26, 987

\bibitem{2008A&A...483..759K} Kovalev, Y.~Y., Lobanov, A.~P., Pushkarev,
A.~B., \& Zensus, J.~A.\ 2008, A\&A, 483, 759 

\bibitem{1998A&A...330...79L} Lobanov, A.~P.\ 1998, A\&A, 330, 79 

\bibitem{Pacholczyk} Pacholczyk A.G. 1970, Radio astrophysics, San~Francisco: Freeman

\bibitem{2005ApJ...621..313V} Virtanen, J.~J.~P., \& Vainio, R.\ 2005, ApJ, 621, 313 

\bibitem{2009ApJ...696L..17K} Kovalev, Y.~Y., et al.\ 2009, ApJL, 696, L17 

\bibitem{Marscher} Marscher A.P. 1983, ApJ, 264, 296

\end{thebibliography}
\end{document}